\documentclass[12pt]{article}
\usepackage{latexsym}
\usepackage{graphicx}
\usepackage[usenames,dvipsnames]{color}

\def\double{\baselineskip 24pt \lineskip 10pt}
\def\be{\begin{equation}}
\def\ee{\end{equation}} 
\def\bea{\begin{eqnarray}}
\def\eea{\end{eqnarray}}
\def\pa{\partial}
\def\sst{\scriptscriptstyle}
\def\mco{\multicolumn}
\def\CPbar{\hbox{{\rm CP}\hskip-1.80em{/}}}%temp replacement due to no font
\def\bu{$\bullet$}
\def\l{\label}
\def\r{\ref}
\def\fn{\footnote}
\def\vc{V^{\frac{2}{3}}}
\def\e{\emph}
\def\rmsd{$\ell_{\textrm{\tiny{rms}}}$ }
\def\rmsdn{$\ell_{\textrm{\tiny{rms}}}$}
\def\nb{$N$-body problem }
\def\nbn{$N$-body problem}
\def\rnb{relational $N$-body problem }
\def\mb{$N$-body }
\def\mbn{$N$-body}

\def\Ra{\Rightarrow}
\def\half{\frac{1}{2}}
\def\nn{\nonumber}
\def\co{\rm{c}}
\def\si{\rm{s}}
\def\wt{\widetilde}
\def\case#1/#2{\textstyle\frac{#1}{#2}}
\def\doublespace{\baselineskip=20pt plus 3pt\message{double space}}
\def\L{\left}
\def\R{right}
\def\noi{\noindent}
\def\vs{\vspace{.2in}}
\def\e{\emph}
\def\noi{\noindent}
\def\id{, i.e., }
\def\p{$\Psi$~}
\def\pn{$\Psi$}

\def\s{${\mathcal S}$ }
\def\sn{${\mathcal S}$}
\def\elb{$I_\textrm{\scriptsize cm}$ }
\def\elbn{$I_\textrm{\scriptsize cm}$}
\def\ccm{$I_\textrm{\scriptsize cm}=0$ }
\def\ccmn{$I_\textrm{\scriptsize cm}=0$}
\def\ecm{$E_\textrm{\scriptsize cm}=0$ }
\def\ecmn{$E_\textrm{\scriptsize cm}=0$}
\def\ec{$E_\textrm{\scriptsize cm}$ }
\def\ecn{$E_\textrm{\scriptsize cm}$}
\def\lcm{${\bf L}_\textrm{\scriptsize cm}=0$ }
\def\lcmn{${\bf L}_\textrm{\scriptsize cm}=0$}
\def\tbe{three-body~}
\def\tben{three-body}
\def\tb{three-body problem }
\def\tbn{three-body problem}
\def\al{Alpha }
\def\aln{Alpha}
\def\pa{Eq.~(\ref{eqm}) }
\def\pan{Eq.~(\ref{eqm})}
\def\el{$\ell_\textrm{\scriptsize rms}$ }
\def\eln{$\ell_\textrm{\scriptsize rms}$}
\def\sc{Schr\"odinger equation }
\def\scn{Schr\"odinger equation}
\def\np{$V_\textrm{\scriptsize Shape}$ }
\def\npn{$V_\textrm{\scriptsize Shape}$}
\def\c{$C$ }
\def\cn{$C$}
\def\isa{$V_\textrm{\scriptsize New}$ }
\def\isan{$V_\textrm{\scriptsize New}$}
\def\po{Poincar\'e }
\def\pon{Poincar\'e}
\def\n{Newtonian }
\def\nn{Newtonian}

\begin{document}

%\begin{center}

%\documentclass[12pt]{article}
%\documentstyle[]{article}
%\usepackage{amssymb}
%\usepackage{epsf}
%\usepackage{latexsym}
%\usepackage{graphicx}
%\usepackage[usenames,dvipsnames]{color}

%\input{psfig}
\bibliographystyle{unstr}
% for BibTex - sorted numerical labels by order of first citation

%****************************ACROS********************************

% Some other useful macros
\def\double{\baselineskip 24pt \lineskip 10pt}
\textheight 8.9 in \textwidth 6.5 in \oddsidemargin -10pt \topmargin
-30pt

\def\be{\begin{equation}}
\def\ee{\end{equation}} 
\def\bea{\begin{eqnarray}}
\def\eea{\end{eqnarray}}
\def\pa{\partial}
\def\sst{\scriptscriptstyle}
\def\mco{\multicolumn}
\def\CPbar{\hbox{{\rm CP}\hskip-1.80em{/}}}%temp replacement due to no font
\def\bu{$\bullet$}
\def\l{\label}
\def\r{\ref}
\def\fn{\footnote}
\def\vc{V^{\frac{2}{3}}}
\def\e{\emph}
\def\rmsd{$\ell_{\textrm{\tiny{rms}}}$ }
\def\rmsdn{$\ell_{\textrm{\tiny{rms}}}$}
\def\nb{$N$-body problem }
\def\nbn{$N$-body problem}
\def\rnb{relational $N$-body problem }
\def\mb{$N$-body }
\def\mbn{$N$-body}

%EQUATION ACROS

\def\Ra{\Rightarrow}
\def\half{\frac{1}{2}}
\def\nn{\nonumber}
\def\co{\rm{c}}
\def\si{\rm{s}}
\def\wt{\widetilde}
\def\case#1/#2{\textstyle\frac{#1}{#2}}
\def\doublespace{\baselineskip=20pt plus 3pt\message{double space}}
\def\L{\left}
\def\R{right}
\def\noi{\noindent}
\def\vs{\vspace{.2in}}
\def\e{\emph}
\def\noi{\noindent}
\def\id{, i.e., }
\def\p{$\Psi$~}
\def\pn{$\Psi$}
\def\s{${\mathcal S}$ }
\def\sn{${\mathcal S}$}
\def\elb{$I_\textrm{\scriptsize cm}$ }
\def\elbn{$I_\textrm{\scriptsize cm}$}
\def\cc{$E_\textrm{\scriptsize cm}$ }
\def\ccn{$E_\textrm{\scriptsize cm}$}
\def\ccm{$I_\textrm{\scriptsize cm}=0$ }
\def\ccmn{$I_\textrm{\scriptsize cm}=0$}
\def\ecm{$E_\textrm{\scriptsize cm}=0$ }
\def\ecmn{$E_\textrm{\scriptsize cm}=0$}
\def\ec{$E_\textrm{\scriptsize cm}$ }
\def\ecn{$E_\textrm{\scriptsize cm}$}
\def\lcm{${\bf L}_\textrm{\scriptsize cm}=0$ }
\def\lcmn{${\bf L}_\textrm{\scriptsize cm}=0$}
\def\com{$C_\textrm{\scriptsize shape}$ }
\def\comn{$C_\textrm{\scriptsize shape}$}
\def\tbe{three-body~}
\def\tben{three-body}
\def\tb{three-body problem }
\def\tbn{three-body problem}
\def\al{Alpha }
\def\aln{Alpha}
\def\pa{Eq.~(\ref{eqm}) }
\def\pan{Eq.~(\ref{eqm})}
\def\el{$\ell_\textrm{\scriptsize rms}$ }
\def\eln{$\ell_\textrm{\scriptsize rms}$}
\def\sc{Schr\"odinger equation }
\def\scn{Schr\"odinger equation}
\def\np{$V_\textrm{\scriptsize Shape}$ }
\def\npn{$V_\textrm{\scriptsize Shape}$}
\def\c{$C_\textrm{\scriptsize shape}$ }
\def\cn{$C_\textrm{\scriptsize shape}$}
\def\isa{$V_\textrm{\scriptsize New}$ }
\def\isan{$V_\textrm{\scriptsize New}$}
\def\po{Poincar\'e }
\def\pon{Poincar\'e}
\def\n{Newtonian }
\def\nn{Newtonian}
\def\no{$N$ }
\def\non{$N$}
\def\eu{Euclidean }
\def\eun{Euclidean}
\def\qm{quantum mechanics }
\def\qmn{quantum mechanics}
\def\wrt{with respect to }
\def\r{$r_{ij}$ }
\def\rn{$r_{ij}$}
\def\ds{$3N-7$ }
\def\dsn{$3N-7$}
\def\dd{$3N-8$ }
\def\ddn{$3N-8$}
\def\me{energy and angular momentum }
\def\men{energy and angular momentum}
\def\hom{A function $f(x_i)$ of multiple arguments $x_i$ is homogeneous of degree $k$ if $f(ax_i)=a^kf(x_i)$, where $a>0$ is a constant.}
\def\mhl{$\ell_\textrm{\scriptsize mhl}$ }
\def\mhln{$\ell_\textrm{\scriptsize mhl}$}
\def\jp{Janus point } 
\def\jpn{Janus point}

%\begin{document}

\title{\bf Complexity and Its Creation}

\author{\large Julian Barbour\footnote{College Farm, South Newington, Banbury, OX15 4JG, UK, email: julian.barbour@physics.ox.ac.uk.}, Zaza Doborjginidze\footnote{Goldsmiths, University of London, MSc Computational Neuroscience, email: zdobo@gold.ac.uk.}, Tim Koslowski\footnote{
Technische Hochschule W\"urzburg-Schweinfurt,
Ignaz-Sch\"on-Stra{\ss}e 11, 97421 Schweinfurt, Germany, email: tim.koslowski@thws.de.}, and Hemant Shukla.\footnote{email: hshuklatmp@yahoo.com.}}

\maketitle

\begin{abstract}
Except for crystalline or random structures, an agreed definition of complexity for intermediate and hence interesting cases does not exist. We fill this gap with a notion of complexity that characterises shapes formed by any finite number of particles greater than or equal to the three needed to define triangle shapes. The resulting shape complexity is a simple scale-invariant quantity that measures the extent to which a collection of particles has a uniform or clustered distribution. As a positive-definite number with an absolute minimum realised on the most uniform distribution the particles can have, it not only characterises all physical structures from crystals to the most complex that can exist but also determines for them a measure that makes richly structured shapes more probable than bland ones. Strikingly, the criterion employed to define the shape complexity forces it to be the product of the two functions that define \n universal gravitation. This suggests both the form and solutions the law of a universe of such particles should have and leads to a theory that not only determines the complexity and probability of any individual shape but also its creation from the maximally uniform shape. It does this moreover in a manner which makes it probable that the cosmological principle, according to which on a sufficiently large scale the universe should have the same appearance everywhere, holds. Our theory relies on universal group-theoretical principles that may allow generalisation to include all forces and general relativity.
\end{abstract}

\section{Introduction\label{int}}

There is extensive literature on the definition of complexity but much less on how complex structures come into existence through physical processes. In the mid 1960s, Kolmogorov's definition of algorithmic complexity \cite{kol} and Solomonoff's theory of inductive inference \cite{sol} did much to define the direction of subsequent research. The development of both approaches relied heavily on Turing's 1950 theory of the eponymous computation machines \cite{tur}, the influence of which can be seen in the definition of Kolmogorov complexity as the length of the shortest programme that generates a given sequence of symbols and in Solomonoff's use of minimal sequences of symbols to define \e{a priori} Bayesian probabilities. 

The influence of Turing's work is seen particularly in a comment of Solomonoff: ``The author feels that all problems in inductive inference, whether they involve continuous or discrete data, or both, can be expressed in the form of the extrapolation of a long sequence of symbols.'' We think that this mindset, perfectly tenable in itself, may nevertheless suffer from two related problems. First, if we are attempting to understand the complexity of physical structures, some protocol must be used to convert the corresponding spatiotemporal observational data into the long sequence of symbols. Rather than clarifying this seems to complicate and possibly even distort the problem of describing and explaining the origin of physical structures as opposed to strings of digits generated by, for example, tossing of coins or dice that are biassed to greater or lesser degree. Second, physics grew out of the attempt to understand the changes in the positions of bodies relative to each other in space. The primary data in this project are positions at a given instant, in the simplest case Cartesian coordinates of particles in \eu space. For this reason, we think positions, rather than a long sequence of symbols, are the appropriate starting point for a definition of complexity. In fact, the greatest human advance in the development of inference came with Newton's second law, which showed how positions and their rates of change at an instant allowed one to make predictions of positions in the future.

In this paper, we therefore begin by developing a theory of complexity on the basis of the positions of a finite number of particles at a given instant. In contrast to Newton, who defined positions of individual particles in absolute space, we define positions of all the particles relative to each other using only the ratios of the distances between them. In the simplest case of three particles, our definition of complexity characterises the shape of the triangle that they form in any instant. Overall scale, which presupposes a ruler outside the universe to determine the size of the triangle, plays no role. Together with the requirement that all particles be treated on the same footing, this suggests a simple and natural expression for the complexity. Strikingly, it is the product of the two functions which represent the irreducible core of Newton's universal theory of gravity that remains when all traces of the absolute elements he introduced to define motion are eliminated. This means that although our definition characterises the complexity of any collection of particles at any instant independently of how it may have come into existence, it nevertheless suggests the form of a law that could govern its evolution---\n gravity. Equally remarkable is the fact that the same requirements which lead to our definition of complexity drastically restrict the solutions for that evolution. The picture that emerges is of a universe that begins with the least possible complexity, and then becomes ever more structured. Thus, we propose a theory of complexity and its creation that has implications on both the smallest and largest scales. We also indicate how our theory might be generalised to match the structure of Einstein's general theory of relativity. At the end of the paper, we make a brief comparison of our approach with some of the leading suggestions made by other authors.

\section{Shape Complexity}
 
Etymologically, something complex is ``formed by a combination of simplest things'' and a complex is ``a whole comprised of interconnected parts'' \cite{oed}. The conceptually simplest things in physics are point particles and spatial separations their simplest interconnections. We define complexity in such terms. As in quantum field theory, fermions can be identified with our particles and force-carrying bosons as the determinants of separations between them in space, taken as an adequate approximation to be \eun. The isotropy of the microwave background provides an observational definition of rest and simultaneity within the universe. Our aim is to present new concepts and possibilities, not their full elaboration. 

We introduce our unifying notion, \e{shape complexity}, through a problem: given a distribution of particles with masses that are ratios of a nominal unit total mass $M=1$,
\be
\sum_i^N m_i=M=1,\label{masses}
\ee 
and Cartesian coordinates ${\bf r}_i$ in space, what simple scale-invariant numbers that take into account all particles on an equal footing can be used to characterise the extent to which they are uniformly distributed or clustered? To be scale-invariant any such number must be homogeneous of degree zero in the distances, the simplest case being 
the ratio of two numbers each with dimension [length]$^1$. Apart from the mass-weighted mean separation, which is seldom used, the simplest candidates that meet our criteria are the \e{root-mean-square length} \eln:
\be
\ell_\textrm{\scriptsize rms}=\sqrt{\sum_{i<j}m_im_jr_{ij}^2},~r_{ij}=|{\bf r}_i-{\bf r}_j|,\label{rms}
\ee
and the \e{mean-harmonic length} \mhln:
\be
\ell_\textrm{\scriptsize mhl}^{-1}=\sum_{i<j}{{m_im_j}\over r_{ij}}.
\ee
Besides their simplicity, each is special. First, up to a constant of no significance that we therefore take to be 1, $\ell_\textrm{\scriptsize rms}^2$ is identically equal to the centre-of-mass moment of inertia \elbn:
\be
\ell_\textrm{\scriptsize rms}^2\equiv I_\textrm{\scriptsize cm}=\sum_i^Nm_i{\bf r}_i^\textrm{\scriptsize cm}\cdot{\bf r}_i^\textrm{\scriptsize cm},\label{icm}
\ee
where ${\bf r}_i^\textrm{\scriptsize cm}$ is the centre-of-mass position of particle $i$. For its part, \mhl is minus the inverse of the Newton potential \isa with $G=1$.\footnote{The Newton potential is generally regarded as a quintessentially physical quantity but appears in our approach as the simplest quantity that, together with \eln, can be used to define complexity.\label{newt}} Thus, our candidate numbers are precisely the tw that characterise \n universal gravitation.

We define the shape complexity as the ratio
\be
C_\textrm{\scriptsize shape}={\ell_\textrm{\scriptsize rms}\over\ell_\textrm{\scriptsize mhl}}\equiv-\ell_\textrm{\scriptsize rms}V_\textrm{\scriptsize New}.\label{sc}
\ee
Our \cn, which in \mb theory is called the shape potential or normalised Newton potential, is a sensitive measure of complexity because close approach or even coincidence of two or a few particles among a sufficient number hardly changes \el but greatly reduces \mhl and increases \c accordingly.\footnote{The choice of \el rather than the mean length enhances this sensitivity since \el gives more weight to large separations and therefore to ratios of them formed with small separations.}

The shape complexity \s is a function on the arena of our approach: \e{shape space} \sn, which consists of all possible (unoriented) shapes defined by the normalised separations 
\be
\bar r_{ij}={r_{ij}\over \ell_\textrm{\scriptsize rms}}\label{bar}
\ee
between \no point particles of given mass ratios.\footnote{{One possible generalisation of our notion of complexity would be to take into account shape orientations.}}

A distinctive aspect of our approach to complexity is that in it discreteness and continuity
play equally important complementary roles. Each of the particles is a discrete entity, while the distances between the particles can change continuously. It is very important that both the masses and distances enter as ratios: relative amount of mass (\ref{masses}) for the particles,\footnote{Besides mass, charge can also play a role and must in the full development of our basic ideas. Although the particles can have different relative masses, all the interesting properties of the complexity stem from the differences of the relative separations (\ref{bar}).} and normalised distances (\ref{bar}) between them. Our belief is that a theory of complexity not based on ratios lacks foundations.\footnote{The first revolution in physics occurred when Copernicus realised that he could, from observations, determine the distances of the planets from the sun as \e{ratios} of the earth--sun distance and from them ratios of orbital speeds. This enabled him to say that, unlike earlier astronomers, he had been able to establish the true form of the universe (by which he meant the solar system). Through the cosmological distance ladder astronomers use Copernicus's insight to establish the true form of the universe as the word is now understood.} The complementary roles of discreteness and continuity together with the exclusive use of ratios distinguish our approach to complexity from many others, especially those mentioned in Sec.~\ref{int} and compared with our approach at the end of the paper. Although they bring a good degree of rigour, mere chronology may have slanted the study of complexity unduly in the direction of computation.\footnote{Turing's paper ``Computing machinery and intelligence'' was published in 1950 \cite{tur}; John McCarthy coined the term artificial intelligence for the Dartmouth Summer Research Project in 1955.} As it happens, our approach is more in the spirit of Turing's study of morphogenesis \cite{turing}, though we start without any pre-existing physical structure. In fact, we doubt if any approach to complexity starts with less than we do.

Our foundational principles, above all the use of ratios, introduce two key properties. The first is that \cn, besides being a measure of complexity, is also the \e{intrinsic size} of the particle distribution it characterises. This follows from the nature of length measurement, in accordance with which, in the ideal case of perfect accuracy, one determines how many units of length $r$ on a ruler fit into the measured length $R$ in the limit $r/R\rightarrow 0$. Now \el is the average of the long separations in a distribution of points, while \mhl is the average of the short ones, so that their ratio, which is \c and a \e{pure number},\footnote{All physical quantities determined experimentally are pure numbers as exemplified by the ratio $r/R$ just given in the text; the difference that \c introduces is the replacement of a single particular unit such as the metre in the metric system by the totality of possible candidates, as in either \el or \mhln. The generalisation from the currently considered case with finite number of candidates to an infinite number will be considered at the end of the paper.} can be said to measure the \e{intrinsic size} of the distribution. No extrinsic ruler is involved. Moreover, a range of sizes that all independent judges can agree on is established; its lower limit is $C_\textrm{\scriptsize shape}^0$, the minimal size the collection of particles can have, but there is no upper limit. In contrast, independent judges could not agree on an analogous range of values of \eln, the value of which (except when zero) presupposes an extrinsic ruler.\footnote{This fact is easily forgotten in connection with expansion of the universe; it corresponds to a change of its shape quantified by intrinsic size, in particular that defined by decrease in the ratio of the typical galactic diameters divided by the typical intergalactic separations; see \cite{jan} for a more detailed discussion.\label{rees}} In contrast, one is presupposed in the determination of \eln, which may be called the \e{extrinsic size} and is purely nominal. 

The second property is that \sn, in being obtained from the \n configuration space ${\mathcal N}$ through quotienting by the similarity group {\sf Sim}, which includes dilatations, is compact. This property of \s has far-reaching consequences. To see this, consider first the \n kinetic metric
\be
\textrm ds^{\textrm{\scriptsize N}}=\sqrt{\sum_im_i\textrm d{\bf r}_i\cdot\textrm d{\bf r}_i},~~
\textrm d{\bf r}_i={\bf r}_i^2-{\bf r}_i^1,\label{my1}
\ee
{as ${\bf r}_i^2$ approaches} ${\bf r}_i^1$. It defines a line element in the Newtonian configuration space ${\mathcal N}$ that reflects not only \e{undeniable intrinsic change} due to change of the shape defined by the particles and expressed through changes in the normalised interparticle separations (\ref{bar}) but also extrinsic change resulting from different relative centre-of-mass positions, orientations, and sizes. They are eliminated by \e{best matching}.\footnote{Best matching is the procedure proposed in \cite{bb} and renamed in \cite{1994} (p. 2862) for finding an \e{intrinsic derivative} free of the effects of \n absolute position and orientation in order to implement Mach's principle; the elimination of absolute scale in \cite{bkm} led on to insights that were wholly unexpected in \cite{bb, 1994} and will be discussed later. } To eliminate the extrinsic contribution to $\textrm ds^{\textrm{\scriptsize N}}$ inherent in the increments $\textrm d{\bf r}_i$ in (\ref{my1}), add to them arbitrary infinitesimal increments generated by {\sf Sim} transformations (translations, rotations, dilations), obtaining trial increments. Since (\ref{my1}) is positive definite and some intrinsic change is assumed, there must exist a `best-matching' set of them that, after division by \elbn, mimimises (\ref{my1}) by bringing the centres of mass to coincidence and reducing to zero the overall rotation and expansion. This leads to a pure number that quantifies infinitesimal \e{difference of shape} and with it a
{\sf Sim}-invariant line element that defines a \e{natural metric} on \sn:
\be
\textrm ds^{\textrm{\scriptsize{\sf Sim}}}_\textrm{\scriptsize bm}={{\textrm{min}}\over{\sf Sim}}\sqrt{{\sum_im_i\textrm d{\bf r}_i\cdot\textrm d{\bf r}_i\over\sum_im_i{\bf r}_i\cdot{\bf r}_i}},~~
\textrm d{\bf r}_i={\bf r}_i^2-{\bf r}_i^1,\label{my}
\ee
The natural metric (\ref{my}){, in which $\sum_im_i{\bf r}_i\cdot{\bf r}_i$ is defined in the centre-of-mass frame,} defines a dimensionless finite volume of any \sn. More significantly, each of the codimension-1 \e{isocomplexity surfaces} into which \s is foliated by \c has a finite natural volume and induced \e{probability measure}, some implications of which have been discussed in \cite{gcc, deut} and others will be important for considerations that follow. 

\section{Properties of the Complexity}

First, we need to say some more things about the complexity. It has a vast number, more than factorial in $N$ and possibly infinite, of critical values (minima or saddles), the shapes of which are known in \mb theory as \e{central configurations} (CCs) (also called relative equilibria, of which, in two dimensions, Saturn's rings and the Lagrange points in the solar system are naturally occurring examples stabilised by centrifugal force).\footnote{For an excellent introduction to CCs see \cite{saari}, in which it is noted that the issue of whether their number is finite or infinite is Steven Smale's sixth problem for mathematics of the 21st century. There is also a very good article on CCs in Wikipedia: Page Version ID: 1219819684.} 

It is easy to find CCs numerically. Since the start of this millennium it has been known that CCs at or very close to the absolute minimum of \c have a remarkably uniform distribution of particles within a sphere with an abrupt boundary. A typical example is shown in Fig.~1. Two things explain its remarkable structure. First, Newton's potential theorem ensures that within a spherically symmetric body the gravitational force at distance $r$ from its centre is equal to the force generated by the mass within the sphere with that radius if concentrated at the origin. Second, \cn, as a product of potentials, generates both attractive gravitational forces and repulsive Hooke forces with strengths precisely right to hold the complete set of particles in relative equilibrium.

\begin{figure}
\begin{center}
\includegraphics[width=0.8\textwidth]{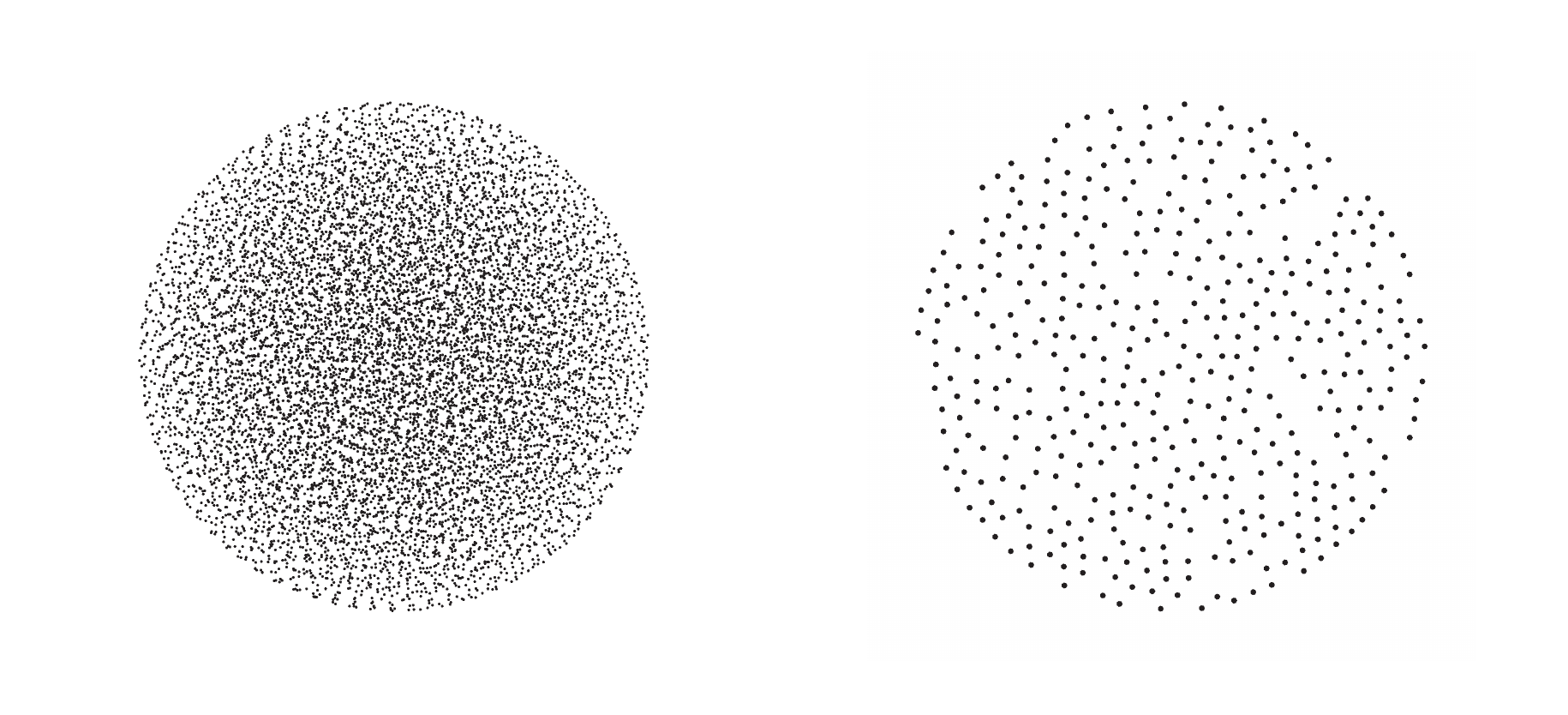}
\caption{\small \e{Left} A 500-particle central configuration with complexity very near the absolute minimum. \e{Right} A section through its centre.}\normalsize\label{fig1}
\end{center}
\end{figure}

\begin{figure}
\begin{center}
\includegraphics[width=0.7\textwidth]{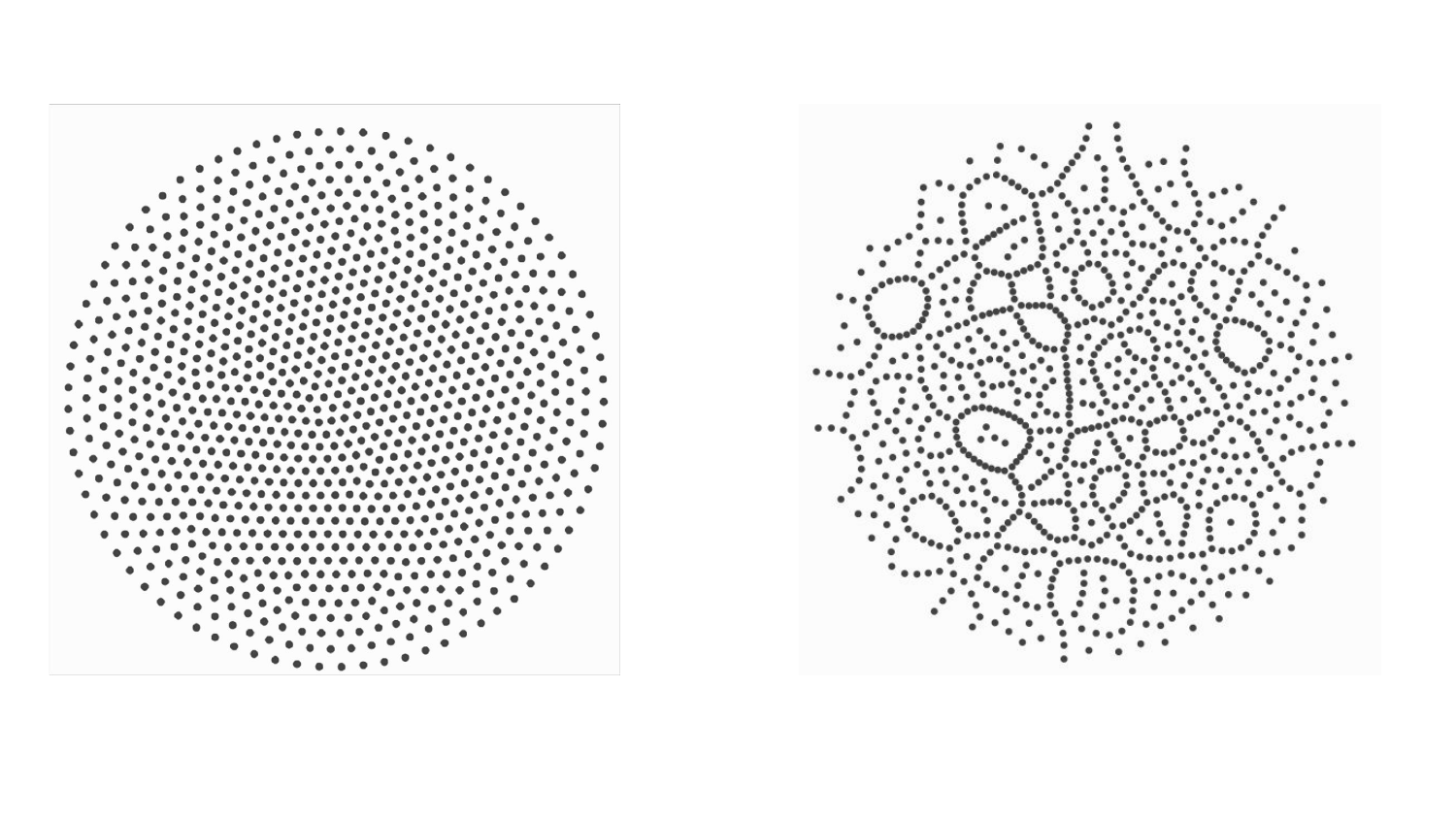}
\caption{\small Two-dimensional 1000-particle central configurations with complexities undetectably above the absolute minimum of \c (left) and about 1.5\% above it.}\normalsize\label{fig2}
\end{center}
\end{figure}

In the spring of 2021, Manuel Izquierdo made a remarkable discovery about the structure of CCs \cite{man}. Hitherto, the main interest in numerical studies had been to find CCs at or at least as close to the absolute minimum of \c as possible and avoid local minima and saddles measurably above it; typical examples found in \cite{bgs} for $N$ up to $10,000$ resembled the one in Fig.~1, while a crystalline BCC-lattice structure in a study with up to $100,000$ particles was found in \cite{japs}. Izquierdo looked initially for CCs in two dimensions and 1000 equal-mass particles. His codes were unlike those used earlier and so did not avoid detection of CCs somewhat above the absolute minimum of \cn. This resulted in his discovery of CCs with the form of the one on the right in Fig.~2, more examples of which are given in \cite{man}. The filaments appear already when \c is little more than 1\% greater than its absolute minimum. Since they clearly consist of interconnected parts made of simplest things they unquestionably manifest complexity in accordance with the etymologies quoted at the start of this section. It will be seen that, unlike the one on the left, which is at or very close to the absolute minimum of \cn, the one on the right, with complexity about $1.5\%$ above the absolute minimum, exhibits a profusion of filamentary structures, closed loops containing a number of particles from $0$ to ten or more, numerous small-scale structures, and particles distributed more or less uniformly within the overall filamentary structures. The increase in the separations between the particles from the centre to the rim of both CCs is due to the calculations having been made in two dimensions, in which Newton's potential theorem does not hold. Izquierdo's calculations for 200 equal-mass particles showed that filaments are also present in three dimensions.\footnote{Jerome Barkley had in fact found such filaments a decade earlier in three dimensions in calculations made for the authors of \cite{bkm} without either they or he recognising them as significant. Barkley's examples, represented stereoscopically and rotating, can be clearly seen by clicking with cursor on the boxes, and if necessary on the images, for 100 particles at \cite{jer}. It will be seen that the number and structural richness of the filaments increases with the complexity.} Twelve two-dimensional ones, found by one of us are shown in Fig.~3 for 100 equal-mass particles (their characterisation by `age' will be justified below). The CC top left is at or very near the absolute minimum of \cn. We hope calculations with supercomputers, with say 100 million particles in three dimensions, can be made to establish just how rich the structure becomes in that case. Meanwhile, Izquierdo's examples in two dimensions are already thought provoking on account of their at least superficial similarity to the structure of the cosmic web \cite{pichon}. 

\begin{figure}
\begin{center}
\includegraphics[width=0.8\textwidth]{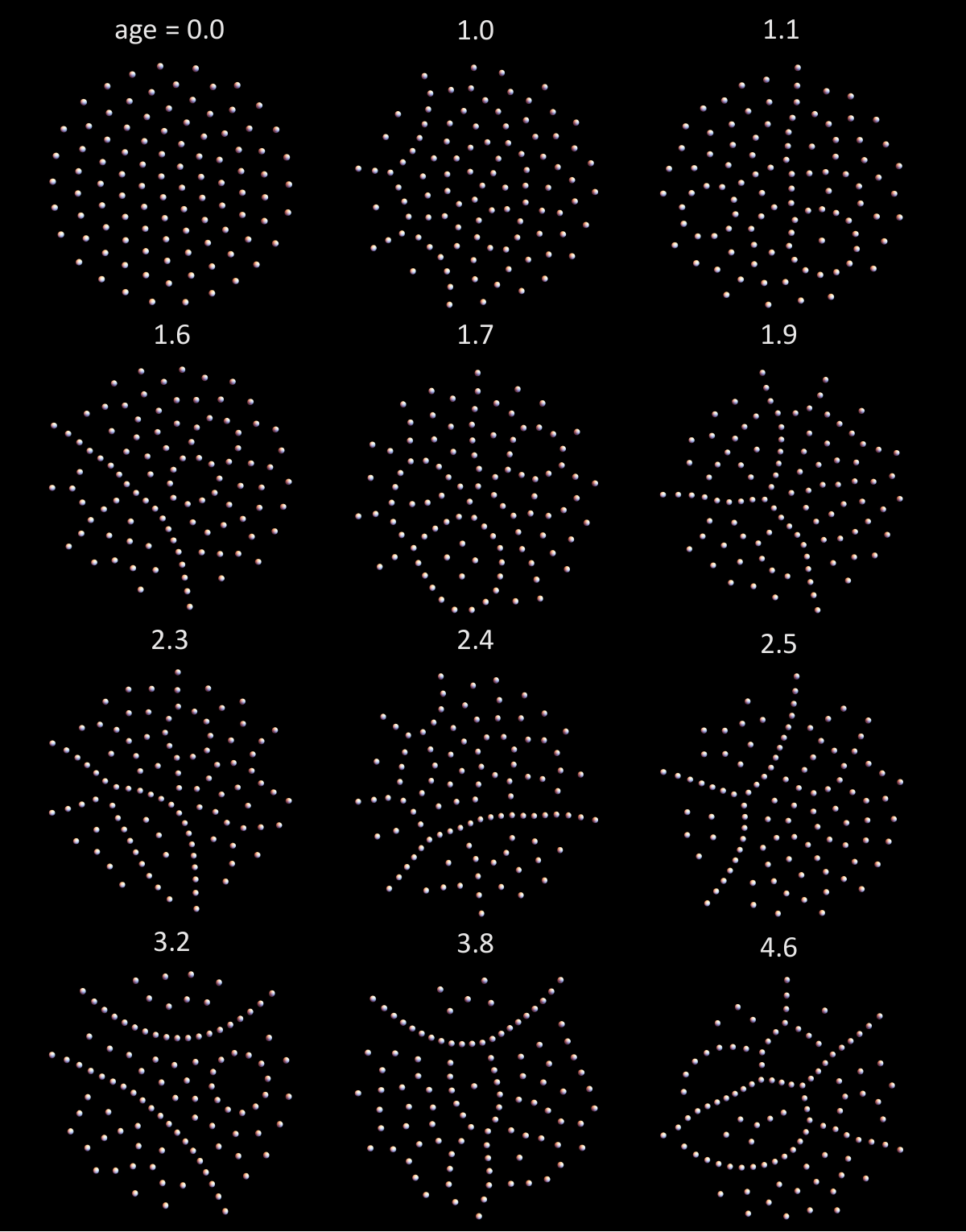}
\caption{\small Twelve central configurations of 100 equal-mass particles. The one top left has \cn$=0.5223$, the next two $0.5258, 0.5261$. We call the difference $0.0035$ of the first two the unit of shape age, say the second has age $1$ and, rounding to the first decimal place, calculate for the others the indicated ages from their \c values.}\normalsize\label{fig1}
\end{center}
\end{figure}

\section{Probabilities of Shapes}

Although shapes with critical values of \cn, which are therefore CCs, are clearly special, there are two reasons why many shapes with the same or nearly the same \c must also look filamentary and indeed as richly structured as the 1000-particle CC on the right in Fig.~2. First, simply because CCs are critical points of \c a relatively large volume of the neighbouring isocomplexity surfaces around any CC defined by the natural metric (\ref{my}) will have very similar, barely perturbed, shapes. Second, a noteworthy generalisation of Boltzmann's $S=k~\textrm{log} ~W$ entropy insight---there are vastly more uniform than non-uniform distributions of molecules in a box---goes a long way to explain the structure of the CCs in Figs.~2 and 3 \cite{OxTalk}. For the particles that define \sn, let $C_\textrm{\scriptsize shape}^\textrm{\scriptsize Alpha}$ be the complexity of shape \e{Alpha} on which the absolute minimum of \c is realised,\footnote{Virtually always unique except for some $N$ smaller than a few tens and providing \s is further quotiented by the permutation group when equal-mass particles are present.} and for any shape $s$ with complexity $C_\textrm{\scriptsize shape}^s>C_\textrm{\scriptsize shape}^\textrm{\scriptsize Alpha}$ define its age $a(s)$ as
\be
a(s)=C_\textrm{\scriptsize shape}^s-C_\textrm{\scriptsize shape}^\textrm{\scriptsize Alpha}.\label{age}
\ee
In accordance with this definition, `older' shapes must be more structured than `younger' ones. Now consider the ways the shape on the left in Fig.~2 can get `older'\id its \c become greater. One obvious way to make this happen is to move a few particles in a localised region of the shape closer to each other. However, there are many more ways to increase the complexity by adjusting all the separations by a small amount than just a few. This ensures two things; first, among the shapes with given \c the overwhelming majority will have shortest separations whose values are sharply peaked. Apart from the slight increase in two dimensions of the separations from the centre to the rim, this is the case for all the CCs in Figs.~2 and 3. In fact, two competing tendencies are at work when \c is increased. On the one hand, the resulting shapes must be more structured but, on the other, Boltzmann's original insight adapted from molecules in a box to this requirement still holds: there are many more ways to create a shape with any given \c greater than the absolute minimum if it contains many more or less equally sized `nuggets' of variety spread uniformly throughout it than large regions of departure from uniformity. This is seen to be the case in the CC on the right in Fig.~2. Nuggets with diameter of about a tenth of the CC, each with its own distinctive structure, are ubiquitous.

This suggests a promising application of our definition of complexity in cosmology, in which the \e{cosmological principle} (CP) is satisfied if for a sufficiently large averaging region the universe has an isotropic and homogeneous matter distribution\id looks the same everywhere. As we have shown, this happens in our CCs. Moreover, it does so in a way that matches the known history of the universe, in which, as it has aged, regions of increasingly greater relative size have been required for the CP to be satisfied. In our CCs, the diameters of `dimes' (to change the metaphor) needed, with the two-dimensional distortion ignored, to cover typical regions are clearly less for the CC on the left in Fig.~2 than for the one on the right. For the 100-particle CCs in Fig.~3, all the CCs except the one top left violate the CP but look much like 100-particle `dime-covered' regions in the CC on the right in Fig.~2. It therefore appears that complexity as we have defined it will give high probabilities to shapes that satisfy the CP. This conclusion is by no means restricted to CCs, since, as we noted, there are vastly more shapes that look very like CCs than there are CCs.\footnote{An interesting, but probably difficult, mathematical problem is the determination of the fraction of any isocomplexity surface in shape space \s that CP-satisfying shapes occupy.}

At the end of the paper we will consider the possibility of fulfilment of the CP in a model universe of infinitely many particles. At this stage, we already note that, besides the essential scale invariance and the special nature of the cofactors in \cn, it is only in three dimensions that the gravitational forces satisfy Newton's potential theorem. This clearly plays a decisive role in creating the conditions under which the cosmological principle is likely to hold with high probability. Indeed, we have already noted that it is satisfied in Figs.~1--3 in a manner that matches qualitatively what has happened in the universe as it has aged. 

We emphasise also that increase of \c has the opposite effect to the entropic `heat-death' of molecules in a box. Inexorable increase of disorder is replaced by its opposite, destructive Boltzmannian uniformisation by \e{creative Boltzmannian uniformisation.} Matching the probability measure defined on each isocomplexity surface, it is structured, not unstructured, shapes that have high probability. As explained in \cite{gcc}, the origin of the difference from the behaviour in thermodynamics and statistical-mechanics is simple: they are the science of systems with phase spaces of bounded Liouville measure. In contrast, \mb solutions exist in an unbounded phase space.

Finally, the filamentary CCs exhibit a striking hierarchical tendency: as the number of particles in the filaments \e{decreases}, the separation between them \e{increases}. In most of the `older' CCs in Fig.~3, three such hierarchies are unambiguously identifiable; the particles not in filaments form a fourth. We do not yet understand this effect, which is redolent of quantum mechanics \cite{deut}, but since just $3N-7$ shape coordinates determine the $N(N-1)/2-1$ normalised inter-particles separations $\bar r_{ij}$ (\ref{bar}) it must have something to do with the unavoidable $\approx N^2$ correlations between the $\bar r_{ij}$, which must all be fitted within a sphere of \cn-dependent intrinsic size. Section 2.9 of \cite{bgs} has many interesting comments about close-packing of spheres as explanation of the shapes associated with the CCs in three dimensions of lowest \cn, but that study pre-dated by two decades the discovery of the filamentary CCs and the intriguing possibilities they suggest. We think generalisation of the close-packing arguments of \cite{bgs}, together with the statistical arguments we have already advanced, could explain the hierarchy of filaments.

\section{Histories}

Having shown that our approach defines not only the complexity of individual $N$-particle shapes but also distinct structures within them, we now point out that, through \n \mb solutions, each shape is uniquely associated with a history of progressive creation of its complexity \c out of the maximal uniformity at Alpha. 
The considerations that have led us to identify these solutions are discussed in \cite{para} and need not be presented here in detail. Our immediate concern is only that the solutions exist, but since they belong to a larger family of solutions whose existence depends critically on central configurations we present details on them in that context.

The existence of these solutions was recognised at the end the 19th century. They are of two kinds and are called, respectively, \e{total-collision} (TC) solutions, in which all the particles collide at their centre of mass, and \e{parabolic-escape} (PE) solutions, in which the shape of the particle distribution `freezes' while the Newtonian scale of the system tends to infinity. In both cases, \mb theoreticians have not been able to find any way to continue the solutions uniquely in accordance with Newton's laws, so that these solutions effectively terminate with extrinsic scales \el$=0$ and \el$=\infty$, respectively. However, what is important from our point of view is that the solutions terminate at central configurations with shapes that are perfectly regular and have intrinsic scales that are neither $0$ nor $\infty$.

Besides the intrinsic vs extrinsic distinction, the time-reversal symmetry of Newton's equations introduces another: we can just as well say the considered solutions \e{begin} rather than terminate at the CCs. In fact, it's what `observers' within them would say and matches the general result of \cite{bkm} that, whenever the energy is non-negative, $E\ge 0$, the behaviour of \c defines arrows of time that are direct consequences of Newton's second law and the homogeneity of degree $-1$ and negative definiteness of \isan. Their existence therefore has nothing to do with entropy increase and growth of disorder. Quite the contrary, they correspond to growth of order, the details of which are given in \cite{bkm,jan}. Moreover, their existence does not require a `past hypothesis' \cite{albert}\id the addition to the existing laws of some special condition in the past. There is such a special condition, but it is a consequence of the laws, not an addition to them. It is manifested in general as a `Janus point': a unique point of minimal \eln, from which, in both time directions, the graph of \el is concave upwards and increases monotonically to infinity. The behaviour of \el therefore defines bidirectional arrows of time. However, \el is an extrinsic scale that `observers' within the \mb universe could not detect. Much more significant is the behaviour of \cn, which increases without bound though with fluctuations in both directions from a minimum near the Janus point. The corresponding arrows are directly visible for internal observers and are aligned with the increase of intrinsic scale\id the observable expansion of the universe.

Arrows of time are manifested most strikingly in TC or PE solutions, in which there is no Janus point and the form of any such solution is like that on one side of a Janus point except that, with growth of \c taken to define the direction of time, they begin with the very special shape of a CC and strongly restricted initial direction away from it (see \cite{chaz, fp}).  Now TC solutions can only exist if the angular momentum $L=0$, but they can have arbitrary energy; the PE solutions can only exist if the energy $E=0$, but they can have arbitrary angular momentum. This raises an interesting question: suppose TC and PE solutions with $E=L=0$ represented in standard \n manner are `projected' to shape space as unparametrised undirected curves. Could observers within them, or mathematicians given only the curves in \sn, say whether they correspond to TC or PE solutions? Or, in other words, could they say in which direction along the curves \el increases? As argued in \cite{para}, we think not. 

However, there is no doubt the succession of shapes in such solutions defines aligned and observable directions of time, increasing order, and growth of intrinsic scale, all of which become ever more pronounced with increasing \non. Within an extended very uniform distribution in which the total energy and angular momentum are both zero, subsystems that become increasingly isolated form with all possible momenta, angular momenta, and energies form and, in general decay, giving rise to new

 We believe such solutions exhibit no trace of the effects of the absolute time, orientation, and scale that Newton introduced when he created dynamics. Their elimination reveals the essence of Newtonian gravity \cite{gcc}: the creation of complexity. Since Newton's theory is in many cases an excellent approximation to Einstein's and shares its group theoretical structure in its Machian solutions with $E=L=0$, we believe there is a good chance this conclusion is also true in general relativity. We may also note that in a big-bang solution in general relativity, proper distance plays the same role as \el in \n theory; this suggests the zero size of the universe in a general-relativistic big bang is a gauge artefact analogous to the use in a \n model universe of unobservable extrinsic size measured by \el as opposed to intrinsic size measured by \cn, which can never be zero.

\section{Comparison with other Definitions}

We conclude with brief comments on some of the better known definitions of complexity.

{\bf Kolomogorov complexity} is the length of the most compressed computer programme that can generate a complete description of a considered object. It assigns higher complexity to random gibberish than text that is intuitively complex, for example, any play by Shakespeare. At any given value of \cn, a tiny fraction of possible shapes will have random (Poisson) distributions like the one on the left in Fig.~4 since such distributions always have a statistically significant number of small separations on a continuum of scales right down arbitrarily near zero. In contrast, in the great majority of all shapes with given \c there will be a large number of nearly equal shortest two-particle separations together with statistical relationships for three-, four-particle separations. Thus, at some \c just above the minimum vastly more shapes will look like the one on the right, which is typical of the distribution of molecules in naturally occurring glasses (or even the sand in deserts from which glass is made). 

{\bf Effective complexity} \cite{gm}, in contrast, places genuine complexity between maximal regularity, as in crystals, and complete randomness. Shape complexity differs from effective complexity in that, for the reason just given, \c characterises the most ordered (crystal) end of its range but effectively excludes all random (Poisson) shapes, which at all values of \c will occur with extremely low probability for large \non. The range of structures that \c characterises is thus closed at one end by minimal complexity but has no upper limit. The universe in its current epoch out to the Hubble radius must have a very high \c because, with fundamental fermions taken as the particles, there is a huge difference between their \el and \mhln. If the currently observed accelerated expansion of the universe continues, this difference can only continue. The objection in \cite{gm} to effective complexity, that it depends on subjective interpretation of experimental data by investigators, is avoided in our approach, which should be applied directly to inter-particle separations (with any observational errors they may contain), not any interpretation that may be given to them.

{\bf Logical depth} \cite{ch} as an approach to measuring the complexity of physical objects faces the problem of finding the smallest Turing machine that could have generated a given object or determining how many steps the machine would take to generate it. This is not a concern for our \c as it it is a single precisely defined scale-invariant number that is readily applied, without the use of Turing machines, to any physical system of interest.

\begin{figure}
\begin{center}
\includegraphics[width=0.7\textwidth]{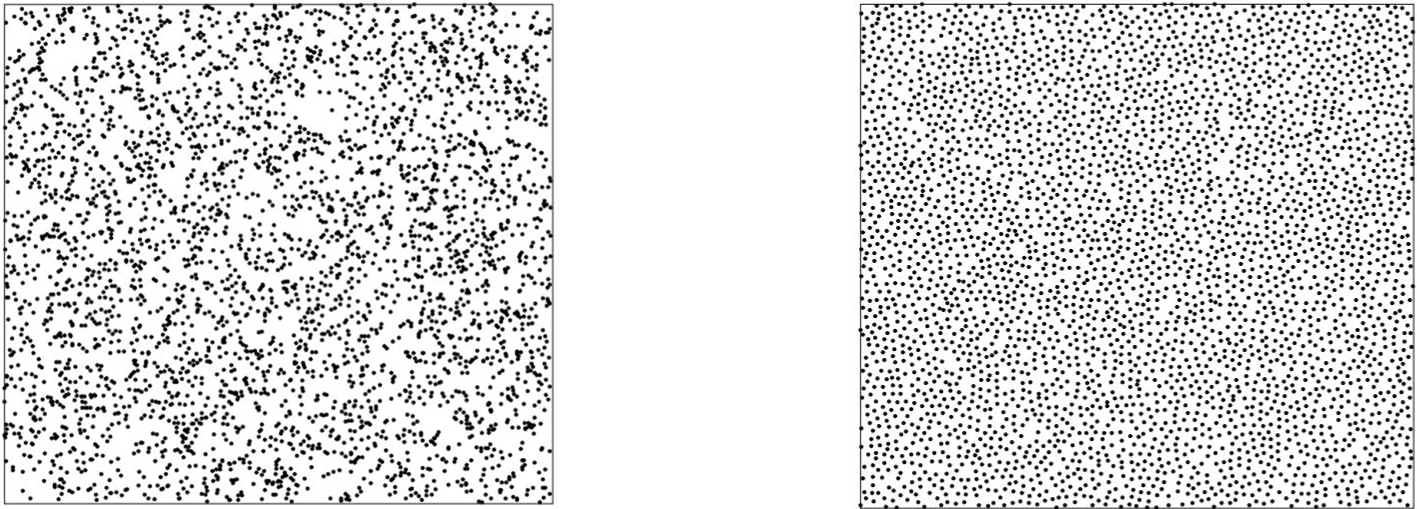}
\caption{\small A random (Poisson) distribution (left) and one qualitatively like a central configuration with \c just above the absolute minimum.}\normalsize\label{fig1}
\end{center}
\end{figure}

{\bf Complexity as thermodynamic depth} \cite{sp} aspires to measure difficulty of creation of structures. When applied to a Hamiltonian system, it ``is equal to the difference between the system's coarse- and fine-grained entropies'' and ``satisfies the requirements that wholly ordered and wholly random systems are not thermodynamically deep.'' The notion of difficulty of creation is attractive in theory particularly if one is sceptical about the idea that every physical system is a digital computer. %Landauer \cite{land} says it is ``one of the remarkably few thrusts in this area that is not conspicuously vacuous''. 
However thermodynamic depth has two significant problems: while \cite{sp} strongly suggest it involves simply coarse-graining the space of microscopic states, it is not clear from the original paper or any of the follow-ups precisely how relevant macroscopic states for which this is to be done are to be selected. In contrast, calculation of \c for any readily distinguished collection of particles is unproblematic as is the `distance' from their Alpha. This distance could be seen as their `difficulty of creation', but our approach removes any real difficulty, which is reduced to the calculation of two complexities.

{\bf Statistical complexity} conceptualises a system as a `message source' and its behaviour as `messages'.    Unlike logical depth, which assumes the use of a universal Turing machine (the most powerful computational model class), statistical complexity relies on a simpler computational model class and has been more successful for measuring real-world phenomena \cite{cy}. However, measuring statistical complexity becomes challenging when the system lacks a straightforward interpretation as a message source.

{\bf Fractal dimension} quantifies the number of copies of a self-similar object at each scale and how the total size of an object changes as the magnification level increases. Ruggedness of fractal-like objects is not the only kind of complexity we find in nature that can and should be measured. The definition is narrow and applicable to only fractal-like objects. A definite advantage of conceptualising in terms of shapes to describe the real world is that the description it gives does not have infinite depth as is the case with mathematically defined fractals, which create self-similar patterns forever. In the real world, the definition of \c ends at the level of individual fermions though, of course, they can be `coarse-grained' into shapes at higher levels, where they could reveal self-similarity. It may also be noted that while fractals and iteratively generated structures like the Sierpinski triangle have striking structures, they necessarily lack the individuality of the images in Fig. 3.

{\bf Hierarchical near-decomposability.} Simon \cite{hs} argued that the most essential common properties of complex systems are hierarchy and near-decomposability, by which is meant that, in hierarchical systems, the number of interactions within each system exceeds those between the systems. We believe this concept, suggested by biological systems, matches our approach well. Consider the hierarchy of the universe out to the Hubble radius, the Galaxy, planetary systems within it, an individual planet, and its satellites. In the standard spacetime representation, they are all bound together by the gravitational field, but it can be `coarse-grained' into Newtonian forces between the centres of mass of the systems at the different levels that are vastly fewer in number than the fermions of the various systems.

{\bf Thanks.} For valuable advice, information, and collaboration (in some cases over many years) that have been critical in his part of this paper, JB thanks, in alphabetical order, Alain Albouy, Alain Chenciner, Pedro Ferreira, Manuel Izquierdo, Tim Koslowski, Flavio Mercati, Richard Montgomery, and David Sloan. JB also thanks Valerie Isham for Maxwell's comment on probability. We all thank Seth Lloyd and Michele Reilly for a helpful discussion of the paper. This work was supported in part by a grant No. 2022-312794 from the Foundational Questions Institute.


\begin{thebibliography}{xxx}

\bibitem{kol} A~Kolmogorov ``On tables of random numbers'' \e{Indian J Statistics, Ser. A}, {\bf 25}, 369 (1963).
\bibitem{sol} R~Solomonoff ``A formal theory of inductive inference'' \e{Information and Control} {\bf 7}, 1 (1964).
\bibitem{tur} A~Turing ``Computing machinery and intelligence'' \e{Mind}, {\bf 49}, 433 (1950).
\bibitem{oed} \e{Online Etymology Dictionary} (2023). 
\bibitem{turing} A~Turing ``The chemical basis of morphogenesis'' \e{Phil. Trans. R. Soc. London B} {\bf 237}, 37 (1952).
\bibitem{jan} J~Barbour \e{The Janus Point} Basic Books (2020), chapter 10.
\bibitem{bb} J~Barbour and B~Bertotti, ``Mach's principle and the structure of dynamical theories'' \e{Proc. R. Soc. Lond.} A {\bf 382}, 295 (1982).
\bibitem{1994} J~Barbour ``The timelessness of quantum gravity: I. The evidence from the classical theory'' \e{Class. Quantum Grav.} {\bf 11}, 2853 (1994). 
\bibitem{bkm} J~Barbour, T~Koslowski, and F~Mercati ``Identification of a gravitational arrow of time''
\e{Phys~Rev~Lett.} \textbf{113}, 181101  (2014); arXiv:1409.0917 [gr-qc].
\bibitem{gcc} J~Barbour ``Gravity's creative core'' arXiv:2301.07657 [gr-qc].
\bibitem{deut} J~Barbour ``Quantum without quantum'' arXiv:2305.13335 [quant-ph], see especially the online talks of [4] of this paper.
\bibitem{saari} D~Saari ``Central configurations---A problem for the 21st century'' [PDF]psu.edu.
\bibitem{man} M~Izquierdo ``Filaments and voids in central configurations'' arXiv:2211.09855.
\bibitem{bgs} R~Battye, G~Gibbons, and P~Sutcliffe ``Central configurations in three dimensions'' arXiv:0201101v2 [hep-th].
\bibitem{japs} H~Totsuji, T~Kishimoto, C~Totsuji, and K~Tsuruta ``Competition between two forms of ordering in finite Coulomb clusters'' \e{Phys. Rev. Lett.}, 125002 (2002).
\bibitem{jer} J~Barkley https//www.nbi.dk/barkley/shapeDynamics/.
\bibitem{para} J~Barbour ``First-principles hints of an alternative cosmological paradigm'' in preparation.
\bibitem{pichon} J~Shim, S~Codi, \e{et al} ``The clustering of critical points in the evolving cosmic web'' \e{MNRAS} {\bf 502}, 3885 (2021); arXiv:2011.04321.
\bibitem{OxTalk} J~Barbour First online talk in [4] of \cite{deut}.
\bibitem{albert} D~Albert \e{Time and Chance} Harvard University Press (2003).
\bibitem{chaz} J~Chazy ``Sur certaines trajectories du probl\`eme des $n$ corps'' \e{Bulletin astronomique} {\bf 34}, 321 (1918).
\bibitem{fp} F~Mercati and P~Reichert ``Total collisions in the \mb shape space'' arXiv:2109.14959 [physics.class-ph].
\bibitem{mv} E~Madurna and A~Venturelli ``Globally minimizing parabolic motions in the \n \mb problem'' \e{Arch. Rational Mech. Anal.} {\bf 194}, 283 (2009); arXiv:1502.06278 [math.DS].
\bibitem{flavio} F~Mercati \e{Shape Dynamics. Relativity and Relationism}, OUP (2018).
\bibitem{max} J~C~Maxwell ``The Scientific Letters and Papers of James Clerk Maxwell'' P~M~Harman, ed., Vol.~1, p.~197.
\bibitem{gm} M~Gell-Mann and S~Lloyd ``Information measures, effective complexity, and total information'' \e{Complexity} {\bf 2}, 44 (1996).
\bibitem{mca}  J McAllister  ``Effective Complexity as a measure of information content'' \e{Phil. Science} {\bf 70}, 302 (2003).
\bibitem{ch} C~Bennett ``Logical depth and physical complexity'' in: \e{The Universal Turing Machine. A Half-Century Review}
\bibitem{ld} ``Logical Depth and Physical Complexity'', in \e{The Universal Turing Machine: A Half-Century Survey}, R~Herken (ed.) Oxford University Press (1998) p. 227.
\bibitem{sp} S~Lloyd and H~Pagels ``Complexity as thermodynamic depth'' \e{Ann. Phys.} {\bf 188}, 186 (1988).
\bibitem{cy} J~Crutchfield and K~Young \e{Phys. Rev. Lett.} {\bf 63}, 105 (1989).
\bibitem{ks} M~von Korff and T~Sander ``Molecular complexity calculated by fractal dimension'' \e{Scientific Reports} {\bf 9}, 967 (2009).
\bibitem{hs} H~Simon ``The architecture of complexity'', \e{Proc. Am. Phil. Soc.} {\bf 106}, 467 (1962).

\end{thebibliography}
\end{document}